\documentclass[aps,reprint,amsfonts,amsmath,amssymb,floatfix]{revtex4-1}
\usepackage{graphicx}
\usepackage{dcolumn}
\usepackage{bm}
\usepackage{enumerate}
\usepackage{color}
\usepackage{parskip}
\usepackage{psfrag}
\usepackage{soul}
\definecolor{amber}{rgb}{1.0, 0.75, 0.0}
\definecolor{redx}{rgb}{1.0, 0.0, 0.5}

\begin{document}

\title{
Hard convex lens-shaped particles: 
Characterization of dense disordered packings 
}

\author{Giorgio Cinacchi}
\email{giorgio.cinacchi@uam.es}
\affiliation{
Departamento de F\'isica Te\'orica de la Materia Condensada,  \\
Instituto de F\'isica  de la Materia Condensada (IFIMAC),    \\
Instituto de Ciencias de Materiales ``Nicol\'as Cabrera'', \\
Universidad Aut\'onoma de Madrid, 
Ciudad Universitaria de Cantoblanco, \\
E-28049 Madrid, Spain
}

\author{Salvatore Torquato}
\email{torquato@princeton.edu}
\affiliation{
Department of Chemistry, 
Department of Physics, \\
Institute for the Science and Technology of Materials, \\
Program for Applied and Computational Mathematics, \\
Princeton University, Princeton, New Jersey 08544, USA
}

\date{\today}

\begin{abstract}
\noindent 
Among the family of hard convex lens-shaped particles (lenses),
the one with aspect ratio equal to 2/3 is `optimal' in the sense that 
the maximally random jammed (MRJ) packings of such lenses achieve 
the highest packing fraction $\phi_{\rm MRJ} \simeq 0.73$.
This value is only a few percent lower than $\phi_{\rm DKP} = 0.76210\dots$,
the packing fraction of the corresponding densest-known crystalline (degenerate) packings. 
By exploiting the appreciably reduced propensity that 
a system of such optimal lenses has to positionally and orientationally order, 
disordered packings of them are progressively generated by
a Monte Carlo method-based procedure
from the dilute equilibrium isotropic fluid phase to the dense nonequilibrium MRJ state.
This allows one to closely monitor how 
the (micro)structure of these packings changes 
in the process of formation of the MRJ state. 
The gradual changes undergone by 
the many structural descriptors calculated
can coherently and consistently be traced back to 
the gradual increase in contacts between the hard particles until
the isostatic mean value of 10 contact neighbors per lens is reached at 
the effectively hyperuniform MRJ state.
Compared to the MRJ state of hard spheres,
the MRJ state of such optimal lenses is 
denser (less porous), more disordered, and rattler-free. 
This set of characteristics makes them good glass formers. 
It is possible that 
this conclusion may also hold for other hard convex uniaxial particles with 
a correspondingly similar aspect ratio, 
be they oblate or prolate, 
and that, 
by using suitable biaxial variants of them,
that set of characteristics might further improve. 
\end{abstract}

\maketitle

\section{introduction}
\label{intro}
One defines a packing as a collection of hard (nonoverlapping) particles in 
a $d$-dimensional Euclidean ($\mathbb{R}^d$) or non-Euclidean space.
Hard-particle packing problems are easy to pose but highly nontrivial to solve.
Indeed, given such a collection of hard particles of a certain shape,
finding the arrangements that maximize the packing fraction $\phi$ is 
a persistent discrete-geometric (optimization) problem 
\cite{discrgeo1,discrgeo2,discrgeo3,discrgeo4,discrgeo5,discrgeo6}
relevant to other sectors of mathematics as well as to science and technology.
In particular, 
hard-particle packing problems naturally emerge whenever 
the subject of the investigation is 
a collection of many particles that mutually interact primarily via 
steeply repulsive interactions irrespective as to whether
their typical length scale is micro- or meso- or macro-scopic.
They are thus pertinent to 
most atomic, molecular, colloidal dense multi-particle systems of interest to  
physics and physical chemistry \cite{crystal,glass,liquid,levine}, 
materials science \cite{callister}, and physico-chemical biology \cite{minton}.

The simplest and most studied among the hard-particle models is 
the one in which the particle shape is a sphere. 
Depending on the specific context and interest, 
packings and systems of hard spheres have been extensively investigated 
from different perspectives and in a variety of situations:
monodisperse and polydisperse, 
equilibrium and nonequilibrium, 
in Euclidean and non-Euclidean spaces across dimensions 
\cite{discrgeo3,discrgeo4,hardspheres1,hardspheres2,hardspheres3,hardspheres4,hardspheres2bis,hardspheres5}.

\begin{figure}
\includegraphics[scale=0.5]{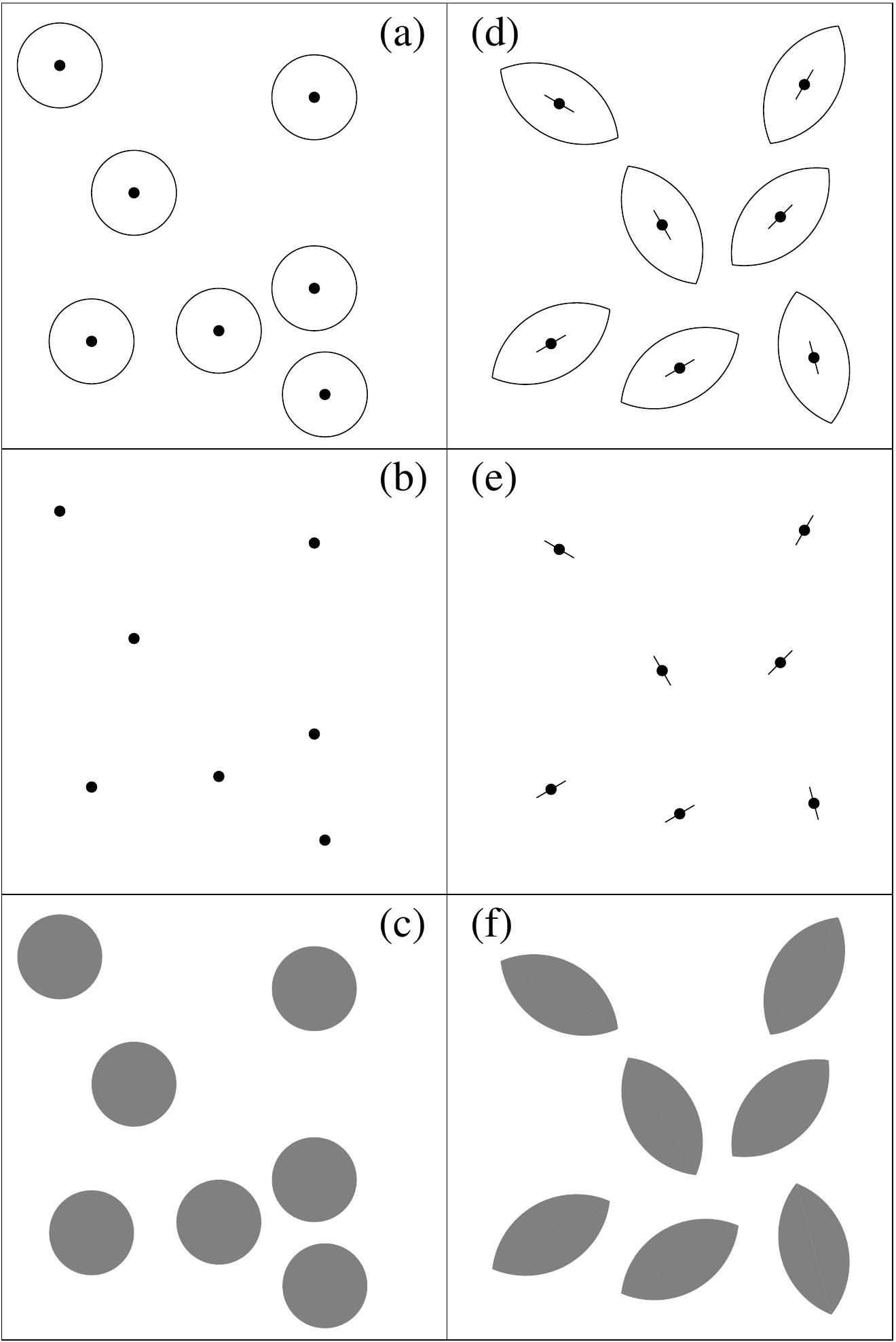}
\caption{ 
Left panels: 
schematic illustration of a packing of hard circles (a) viewed 
as a pattern of points formed by the centers (b) and 
as a two-phase medium with 
the matrix phase being the white region and the particle phase being the gray region (c).
Right panels: 
schematic illustration of a packing of hard almond-shaped particles (d) viewed 
as a pattern of points formed by the centroids 
each one associated with a unit vector along the respective particle symmetry axis (e) and 
as a two-phase medium with 
the matrix phase being the white region and the particle phase being the gray region (f).
}
\label{figillustra}
\end{figure}

In the course of most of these studies,
the structural characterization of 
the hard-sphere packings [Fig. \ref{figillustra} (a)] 
has amounted to 
the structural characterization of 
the patterns of points formed by their centers [Fig. \ref{figillustra} (b)].
This involves the calculation of suitable positional and bond-orientational correlation functions. 
On many other occasions, 
hard-sphere packings have also been viewed as two-phase media,
with 
the hard-particle exterior constituting the matrix phase ${\cal V}_1$ and
the complementary union of the hard-particle interiors constituting the particle phase ${\cal V}_2$ [Fig. \ref{figillustra} (c)] \cite{bookhetmat}.
The corresponding structural characterization involves the calculation of 
a sequence of positional $n$-point probability functions as well as 
a pore-size distribution function \cite{bookhetmat}.
These functions can then lead to an estimate of 
the effective electromagnetic, mechanical, and transport properties of 
a heterogeneous material made of phase ${\cal V}_1$ and phase ${\cal V}_2$ \cite{bookhetmat}.

More recently, the hard-sphere model has been extended to study 
dense packings and systems of hard nonspherical particles, 
which introduce rotational degrees of freedom 
\cite{hardspheres4,hardspheres5}. 
Examples of nonspherical shapes examined include 
ellipsoids \cite{ellipsoids1,ellipsoids2,ellipsoids3,ellipsoids4,ellipsoids5,ellipsoids6,ellipsoids7,ellfurther,ellipsoids8,ellipsoids9}, 
spherocylinders \cite{spherocylinders1,spherocylinders2,spherocylinders3,spherocylinders4,spherocyl,spherocylfurther},
cutspheres \cite{cutspheres1,cutspheres2}, 
superballs \cite{superballs1,superballs2} and
polyhedra \cite{polyhedra1,polyhedra2,polyhedra3,polyhedra4,polyhedra5,polyhedra6,polyhedra7,polyhedra7,polyhedra8,escobedo,truncatetra,glotzer}.
Characterizing the structure and physical properties of 
equilibrium and nonequilbrium states of 
dense packings and systems of hard nonspherical particles continues 
to present many fascinating challenges \cite{hardspheres4,hardspheres5}.

We previously investigated 
the densest-known (crystalline) packings, equilibrium phase behavior, and nonequilibrium jammed states of 
hard convex lens-shaped particles (lenses) \cite{lens1,lens2}.
These hard $\mathsf{{D}_{\infty h}}$-symmetric discoidal particles correspond to 
the intersection volume of two congruent three-dimensional spheres.
By varying the radius of or the center-to-center distance between these spheres,
the class of lenses can be generated. 
Each member of this class is identified by the aspect ratio $\kappa = b/a$, 
with $a$ one of the infinite ${\mathsf{C}_{2}}$ axes and $b$ the ${{\mathsf{C_{\infty}}}}$ axis.
The lens shape interpolates between
the hard infinitesimally-thin disc ($\kappa=0$) and 
the hard-sphere ($\kappa=1$) models.  
\begin{figure}
\psfrag{aaa}{$a$}
\psfrag{bbb}{$b$}
\includegraphics[scale=0.67]{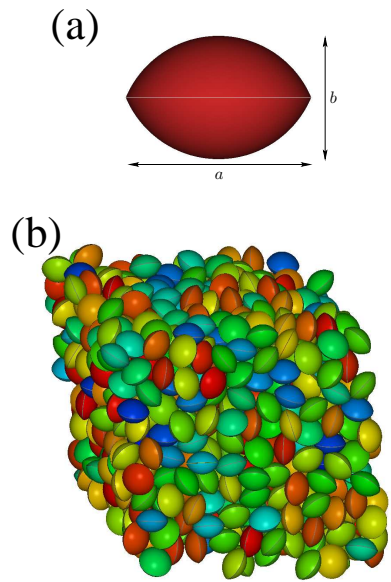}
\caption{
(a) image of a lens with $\kappa = b/a = 2/3$; 
(b) image of a MRJ packing of lenses with $\kappa =2/3$ with 
particles colored according to the angle that 
their $\mathsf{C}_{\infty}$ axis makes with an axis of the laboratory reference frame: 
the cooler the color of a particle 
the smaller the angle that its axis forms with that axis od the laboratory reference frame.
}
\label{figura1}
\end{figure}

This work reports on 
the characterization of 
the (micro)structure of 
monodisperse (positionally and orientationally) disordered packings of 
lenses with $\kappa=2/3$  
henceforth designated as `optimal' lenses [Fig. \ref{figura1} (a)].
This distinguished case is designated in this manner for two intertwined reasons:
\begin{enumerate}[(I)] 
\item 
Systems of optimal lenses have 
a substantially reduced propensity to 
positionally and orientationally order \cite{lens2}.
Disordered packings can then be generated by 
gently compressing from 
their equilibrium isotropic fluid phase up to 
the nonequilibrium maximally random jammed (MRJ) state. 
This state is the one among all strictly jammed \cite{hardspheres4,hardspheres5} states 
that minimizes suitably defined order metrics \cite{mrj}.
Such a gentle compression allows one 
to closely monitor the formation of this special hard-particle state.
In addition to an extremal packing fraction $\phi_{\rm MRJ}$ and 
an isostatic mean number of contacts per particle $\mathtt{Z}=2d_{\rm f}$,
where $d_{\rm f}$ is the number of degrees of freedom for a single particle 
\cite{hardspheres4,hardspheres5},
a hard-particle MRJ state has the particularly important attribute of (effective) hyperuniformity 
\cite{hyperuniformity1,hyperuniformity2,hyperuniformity3}.
Hyperuniformity is a global property of a system that involves 
an anomalous suppression of density fluctuations at large length scales, 
which is completely accessible via scattering in the infinite-wavelength limit 
\cite{hyperuniformity1,hyperuniformity2,hyperuniformity3}.
This unusual characteristic that special disordered systems, 
including hard-particle MRJ states \cite{MRJhyper1,MRJhyper2,MRJhyper3}, 
possess 
is shared with perfect crystals and quasicrystals \cite{hyperuniformity1,hyperuniformity2,hyperuniformity3}.
\item   
The MRJ state of lenses with $\kappa \simeq 2/3$ [Fig. \ref{figura1} (b)] is 
the one most densely packed:
the graph of $\phi_{\rm MRJ}$ versus $\kappa$ has 
its absolute maximum at $\kappa \simeq 2/3$ \cite{lens2}.
This maximal value of $\phi_{\rm MRJ} \simeq 0.73$ is 
only $\simeq 4$\% smaller than $\phi_{\rm DKP} = 0.76210\dots$,
the packing fraction of the densest-known 
(positionally and orientationally ordered, i.e., crystalline as well as degenerate) packings of 
lenses with the same value of $\kappa$ \cite{lens1,lens2}.
It is also very close to 
the packing fractions reached by jammed packings of lenses with different values of $\kappa$ 
in which positional (plastic-crystalline) or orientational (nematic liquid-crystalline) order are 
however present \cite{lens2}.
This fact suggests a strict interrelationship between
the propensity to form (plastic- or liquid-crystalline) mesophases and 
the capability of reaching very dense jammed states without 
the need of introducing positional or orientational order \cite{lens2}.
\end{enumerate}     

In analogy with hard-sphere packings, 
the characterization of 
the (micro)structure of packings of hard nonspherical particles, such as lenses, can be 
simplified by viewing them as patterns of points formed by their centroids.
This is, however, insufficient:
the nonsphericity necessarily leads to associating the position of any centroid
to a set of variables defining the orientation of the corresponding particle.
In case of $\mathsf{{D}_{\infty h}}$-symmetric particles, such as lenses,
this set is formed by the two Euler angles defining 
the orientation of the unit vector along the particle ${{\mathsf{C_{\infty}}}}$ axis.
Thus, packings of hard $\mathsf{{D}_{\infty h}}$-symmetric particles, such as lenses [Fig. \ref{figillustra} (d)], 
can actually be viewed as 
patterns of points where each one is associated with a unit vector [Fig. \ref{figillustra} (e)] 
rather than patterns of sole points.
Consequently, their structural characterisation involves 
not only the calculation of 
suitable positional and bond-orientational correlation functions, 
but also orientational correlation functions.
Naturally, packings of hard nonspherical particles, such as lenses, can also be viewed 
as two-phase media [Fig. \ref{figillustra} (f)]. 
Their structural characterisation involves the calculation of 
the same sequence of $n$-point probability functions as well as 
the pore-size distribution function \cite{bookhetmat}.

By calculating a number of structural descriptors, 
many disordered packings of optimal lenses,
generated by a Monte Carlo method-based procedure 
from the dilute equilibrium isotropic fluid phase up to the dense nonequilibrium MRJ state, 
are characterized.
Similarly to the hard-sphere MRJ state,
the MRJ state of optimal lenses is found to be isostatic and (effectively) hyperuniform 
but, compared to the former, the latter is
denser (less porous), more disordered  and rattler-free.
Thus, even though monodisperse, optimal lenses promise to be 
very good (positional and orientational) glass formers.

The rest of this work consists of the following four Sections:
Section \ref{packchar}, 
that lists all the quantities that have been
calculated to statistically describe the (micro)structure of optimal-lens packings; 
Section \ref{packgen}, 
that very briefly recalls how 
these optimal-lens positionally and orientationally disordered packings have been generated via 
a simple Monte Carlo method-based procedure;
Section \ref{results}, that presents all the results;
Section \ref{conclu}, that makes a few concluding comments.

\section{lens packing (micro)structure characterization}
\label{packchar}

In the characterization of their (micro)structure,
the packings were viewed  
either as  
patterns of the $N$ lens centroids
$\displaystyle \{ \mathbf{r}_1, \cdots, \mathbf{r}_i, \cdots, \mathbf{r}_N \}$ 
each one associated with the respective unit vector along the lens $\mathsf{{C}_{\infty}}$ axis
$\displaystyle \{ \hat{\mathbf{u}}_1, \cdots, \hat{\mathbf{u}}_i, \cdots, \hat{\mathbf{u}}_N \}$ 
or 
as two-phase media 
with the lens exterior constituting the matrix phase ${\mathcal{V}}_1$ and 
the complementary union of the lens interiors constituting the particle phase ${\mathcal{V}}_2$. 

\subsection{
Real-space pair correlation functions and reciprocal-space structure factor
}

If lens packings are viewed as patterns of points each one associated with a unit vector,
their (micro)structure can be characterized by 
several real-space positional, orientational and bond-orientational pair correlation functions 
along with the reciprocal-space structure factor. 

The set of real-space (real-distance) pair correlation functions includes 
$g(r)$, 
the most basic positional pair correlation function,
proportional to 
the conditional probability density of finding the centroid of a lens $j$
at a distance $r$ from the centroid of a lens $i$ 
\cite{liquid,hardspheres1,hardspheres2,hardspheres3,hardspheres4,hardspheres2bis,hardspheres5,bookhetmat}, 
along with 
the orientational pair correlation functions
${\mathcal{G}}_{2n}^{\hat{\mathbf{u}}\hat{\mathbf{u}}} (r)$
and 
the bond-orientational pair correlation functions
${\mathcal{G}}_{2n}^{\hat{\mathbf{u}}\hat{\mathbf{r}}} (r)$.
The latter functions are
respectively defined as:
\begin{equation}
\mathcal{G}_{2n}^{\hat{\mathbf{u}}\hat{\mathbf{u}}} (r) = 
\left \langle 
\frac
{\sum_{i=1}^N \sum_{i \ne j}^N P_{2n} 
\left( \hat{\mathbf{u}}_i \cdot \hat{\mathbf{u}}_j  \right)
\delta \left(r - r_{ij} \right)}
{\sum_{i=1}^N \sum_{i\ne j}^N 
\delta \left(r - r_{ij} \right)} \right \rangle
\label{gruppo1}
\end{equation}
and 
\begin{equation}
\mathcal{G}_{2n}^{\hat{\mathbf{u}}\hat{\mathbf{r}}} (r) = 
\left \langle 
\frac
{\sum_{i=1}^N \sum_{i \ne j}^N P_{2n} 
\left( \hat{\mathbf{u}}_i \cdot \hat{\mathbf{r}}_{ij}  \right)
\delta \left(r - r_{ij} \right)}
{\sum_{i=1}^N \sum_{i\ne j}^N 
\delta \left(r - r_{ij} \right)} \right \rangle
\label{gruppo2}
\end{equation}
with: $r_{ij} = \left|\mathbf{r}_{ij}\right|=\left|\mathbf{r}_j-\mathbf{r}_i\right|$, 
$\hat{\mathbf{r}}_{ij} = \mathbf{r}_{ij}/{r}_{ij}$;
$\delta (r)$ is the radial Dirac $\delta$ function,
$P_{m} (x)$ the $m$-order Legendre polynomial and 
angular brackets indicate an average over configurations.
The $\mathcal{G}_{2n}^{\hat{\mathbf{u}}\hat{\mathbf{u}}} (r)$'s
measure the degree of correlation in the orientations of two lenses whose
centroids are separated by a distance $r$.  
The $\mathcal{G}_{2n}^{\hat{\mathbf{u}}\hat{\mathbf{r}}} (r)$'s
measure the degree of orientational order of
the fictitious `bond' $ \mathbf{r}_{ij} $,
established between the centroids of two lenses $i$ and $j$, 
with respect to $\hat{\mathbf{u}}_i$.

Together with these real-space pair correlation functions, 
the orientationally averaged structure factor ${\mathcal{S}}(k)$,
essentially the Fourier transform of $h(r)=g(r)-1$ \cite{liquid,hardspheres1,hardspheres4,hardspheres5,hyperuniformity1,hyperuniformity3},
was also calculated.
${\mathcal{S}}(k)$ is defined as:
\begin{equation}
{\cal S}(k)=\frac{1}{N} \left \langle \left | \sum_{j=1}^N \mathsf{e}^{i \mathbf{k} \cdot {\mathbf{{r}}}_j}\right|^2 \right \rangle_{\widehat{\mathbf{k}}}, 
\mathbf{k} \neq \mathbf{0} 
\label{avestruct}
\end{equation}
with $k =|\mathbf{k}|$,  
$\mathbf{k}$  a reciprocal-space vector 
and the symbol 
$\displaystyle \langle \rangle_{\widehat{\mathbf{k}}}$
indicating an average over 
the reciprocal-space vectors sharing the same value of $k$ as well as over configurations.
The calculation of ${\mathcal{S}}(k)$, 
made directly according to Eq. \ref{avestruct} rather than Fourier transforming $h(r)$, 
is important because its value in the limit $k \rightarrow 0$ informs 
one about 
the degree of hyperuniformity of a system.
In fact, a hyperuniform many-particle system 
in $\mathbb{R}^d$ is one in which ${\mathcal S}(k)$ tends to zero in the limit $k\to 0$ 
\cite{hyperuniformity1,hyperuniformity2,hyperuniformity3}.
Equivalently, it is one in which
the local number variance associated with a spherical window of radius $R$, 
scaled by $R^d$, vanishes in the large-$R$ limit
\cite{hyperuniformity1,hyperuniformity2,hyperuniformity3}.

\subsection{Pair correlation function of the scaled distance and contact statistics}

Given the hard, convex and nonspherical character of the particles constituting the packings, 
it is useful to define a positional pair correlation function ${\mathtt{g}}(s)$ of 
the scaled distance
$s_{ij}=r_{ij}/{{\cal D}}\left({\hat{\mathbf{r}}}_{ij}, {\hat{\mathbf{u}}}_i, {\hat{\mathbf{u}}}_j \right)$,
with ${\cal D}\left({\hat{\mathbf{r}}}_{ij}, {\hat{\mathbf{u}}}_i, {\hat{\mathbf{u}}}_j \right)$
the distance of closest approach or contact distance 
between lens $i$ and lens $j$.
One way to define ${\mathtt{g}}(s)$ is 
to mimic the most basic physical interpretation of $g(r)$ as 
the ratio between the mean number of centroids found in 
a spherical shell of radii $r$ and $r+dr$ centred on a given centroid and 
the mean number of such centroids in a isodense ideal gas (Poissonian point pattern):
\begin{equation}
{\mathtt{g}}(s)= \frac{\left \langle d n(s) \right \rangle}{\left \langle d n_{\rm id}(s) \right \rangle}    = 
\frac{\left \langle d n(s) \right \rangle}
{3 \varrho \left \langle
v_{exc} \right \rangle s^2 ds},
\end{equation}
with $ \left \langle d n(s) \right \rangle $ the number of centroids having 
a scaled distance from a central centroid $\in [s, s+ds]$ averaged over
central centroids and configurations and
$ \left \langle d n_{\rm id}(s) \right \rangle$ 
the analogous mean number of centroids in a isodense ideal gas;
in its turn, $\left \langle d n_{\rm id}(s) \right \rangle$ is given by 
$\displaystyle 3 \varrho \left \langle
v_{exc} \right \rangle s^2 ds$, 
with 
$\varrho$ the number density and $\left \langle v_{exc} \right \rangle$ 
the expected excluded volume associated with one lens 
averaged over $\hat{\mathbf{u}}_i$ and $\hat{\mathbf{u}}_j$ \cite{rostock}.
Differently than
the full many-variable pair correlation function 
$g(\mathbf{r}_{ij},\hat{\mathbf{u}}_i, \hat{\mathbf{u}}_j)$, 
the positional pair correlation function ${\mathtt{g}}(s)$
can be more directly compared to 
$g(r)$ of a hard-sphere system, $g_{\rm hs}(r)$, and 
its value in the 
$\lim_{s \rightarrow {1^+}} {\mathtt{g}}(s) = {\mathtt{g}}(1^+)$ 
is analogously 
related to the pressure $P$ of 
a statistically homogeneous and isotropic system:
\begin{equation}
\beta P = \varrho \left[ 1 + \frac{1}{2} \varrho \left\langle\left\langle
v_{exc} \right \rangle \right \rangle \mathtt{g} (1^+)  \right], 
\end{equation}
with $\beta=1/(k_BT)$ and $k_B$ the Boltzmann constant and $T$ the absolute temperature. 
For a lens in a configuration, 
the occupacy of the bin at $s=1$ defines the number, 
$\displaystyle {\mathtt{n}}_{\rm c}$, of lenses at contact with it. 
One can then calculate the probability density that 
a lens has ${\mathtt{n}}_{\rm c}$ contact neighbors, 
$\Pi \left( {\mathtt{n}}_{\rm c}  \right)$, 
along with its first moment, 
the mean value of $\displaystyle {\mathtt{n}}_{\rm c}$, as a function of $\phi$, 
$\left \langle {\mathtt{n}}_{\rm c} \right \rangle (\phi)={\mathtt{Z}}(\phi)$. 

\subsection{Lens packings as two-phase media}

It is useful to view lens packings as two-phase media,
in which phase 1 (matrix phase) comprises the space exterior to the particles, ${\cal V}_1$,
and phase 2 (particle phase) comprises the space occupied by the particles, ${\cal V}_2$,
such that 
${\cal V}_1 \cup {\cal V}_2 = \mathbb{R}^3$.
Their (micro)structure can then be characterized by an infinite hierarchy of
$n$-point probability functions \cite{bookhetmat}.
These functions are defined in terms of the phase indicator function:
\begin{eqnarray}
I(\mathbf{x})=\left\{ 
\begin{array}{lr}
0 & : \mathbf{x} \in {{\mathcal{V}}_1} \\
1 & : \mathbf{x} \in {{\mathcal{V}}_2}
\end{array}
\right.
\end{eqnarray}
with $\mathbf{x} \in V \subset  {\mathbb{R}}^3$
as:
\begin{equation}
S_{n} \left( \mathbf{x}_1, ..., \mathbf{x}_n \right) =
\left \langle I(\mathbf{x}_1) ... I(\mathbf{x}_n) \right \rangle.
\end{equation}
This $n$-point function
is the probability of finding $n$ randomly selected points
at positions ${\bf x}_1, \ldots, {\bf x}_n$ in phase 2.
For statistically homogeneous media, 
the one-point function is simply equal to 
the packing fraction, i.e., $S_1({\mathbf{x}}) \equiv \phi$, and 
the two-point function depends only on the displacement vector, 
$S_2({\mathbf x}_1,{\mathbf x}_2)=S_2({\mathbf x}_2-{\mathbf x}_1)$.
If the system is also statistically isotropic, 
the two-point function depends only on 
the modulus of the distance between the two points: 
$S_2(\mathbf{x}_1,\mathbf{x}_2)=S_2(\left|\mathbf{x}_2-\mathbf{x}_1\right|)=S_2(x)$. 
Furthermore, statistical homogeneity suffices to allow
$S_2(x)$ to be separated into
an `internal' component,
$S_{2_{\rm{int}}} (x)$, 
that gives the probability that 
the two randomly 
selected 
points will be at a distance $x$ and lie inside the same particle,
and an `external' component, 
$S_{2_{\rm{ext}}} (x)$,
that gives the probability that 
the two randomly 
selected 
points will be at a distance $x$ and lie inside two different particles:
\begin{equation}
S_2 (x) = S_{2_{\rm{int}}} (x) + S_{2_{\rm{ext}}} (x).
\end{equation}
While the former component is a single-particle quantity that 
does not depend on $\phi$ except for a multiplicative factor, 
it is the latter component that contains information on 
how the (micro)structure of the packings changes with $\phi$ via
pair correlations.
Then, it is convenient to write $S_2(x)$ 
in terms of the positional pair correlation functions 
$\Sigma_{2_{\rm{int}}}(x)$ and $\Sigma_{2_{\rm{ext}}}(x)$:
\begin{equation}
S_2(x)=\phi^2 \left[ \Sigma_{2_{\rm{int}}}(x) + \Sigma_{2_{\rm{ext}}}(x) \right].
\label{S2toSigma}
\end{equation}
The one-body term
$\Sigma_{2_{\rm{int}}}(x) = \phi^{-2} S_{2_{\rm{int}}}(x)$ 
is calculable once for each particle type and 
only contains information about
the particle shape and size.
The two-body term
$\Sigma_{2_{\rm{ext}}}(x) = \phi^{-2} S_{2_{\rm{ext}}}(x)$  
more importantly contains pair correlation information. 
Then, it is natural to introduce the autocovariance function $\chi(x)$ \cite{bookhetmat}:
\begin{equation}
\chi(x) = S_2(x) - \phi^2 = 
\phi^2 \left[ \Sigma_{2_{\rm{int}}}(x) + \Sigma_{2_{\rm{ext}}}(x) - 1 \right  ].
\end{equation}

The Fourier transform of $\chi(x)$ defines 
the spectral density $\hat{\chi} (k)$ \cite{bookhetmat},
which is analogously expressable as the sum of two components,
$\hat{\chi}_{{\rm{int}}} (k)$ and $\hat{\chi}_{{\rm{ext}}} (k)$:
\begin{equation}
\hat{\chi} (k)= \hat{\chi}_{{\rm{int}}} (k) + \hat{\chi}_{{\rm{ext}}} (k),
\label{spectradens1}
\end{equation}
The internal component is given by:
\begin{equation}
\hat{\chi}_{{\rm{int}}} (k) = 
\frac{\varrho}{4\pi} \int d \hat{\mathbf{k}} \left [ \hat{m}(\mathbf{k})\right]^2,
\label{spectradens2}
\end{equation}
where
$\hat{m}(\mathbf{k})$ is the Fourier transform of
the single-particle indicator function \cite{bookhetmat} 
\begin{eqnarray}
m(\mathbf{x})=\left\{ 
\begin{array}{lr}
0 & : \mathbf{x} \not \in {\rm {particle}}  \\
1 & : \mathbf{x}      \in {\rm {particle}}
\end{array}
\right.
.
\end{eqnarray}
The external component is given by:
\begin{equation}
\hat{\chi}_{{\rm{ext}}} (k) = \phi^2 \frac{4\pi}{k}  
\int_0^{\infty} dx ~ x ~ \sin(k x) \left[\Sigma_{2_{\rm{ext}}} (x) -1\right]. 
\label{spectradens3}
\end{equation}
In analogy with what occurs with $h(r)$ and ${\mathcal{S}}(k)$, 
knowledge of $\hat{m}(\mathbf{k})$ and $\Sigma_{2_{\rm{ext}}} (x)$
would allow one to calculate $\hat{\chi} (k)$ via 
Eqs. \ref{spectradens1},\ref{spectradens2},\ref{spectradens3}.
In analogy to ${\mathcal{S}}(k)$, 
${\hat{\chi}}_2(k)$ was  
directly calculated using \cite{bookhetmat}:
\begin{equation}
{\hat{\chi}}(k)=\frac{1}{V} \left \langle 
\left | \sum_{j=1}^N \hat{m}_j(\mathbf{k}) 
\mathsf{e}^{i \mathbf{k} \cdot {\mathbf{{r}}_j}}\right|^2 
\right \rangle_{\widehat{\mathbf{k}}}, 
\mathbf{k} \neq \mathbf{0}.
\end{equation}
The calculation of ${\hat{\chi}}(k)$ is important because 
its value in the limit $k\rightarrow 0$ 
informs one about
the degree of hyperuniformity for a two-phase medium. 
In fact, a hyperuniform two-phase system 
in $\mathbb{R}^d$ is one in 
which ${\tilde \chi}(k)$ tends to zero
in the limit $k\to 0$ 
\cite{hyperuniformity2,hyperuniformity3}.
Equivalently, it is one in which
the local volume fraction variance associated with a spherical
window of radius $R$, scaled by $R^d$, vanishes
in the large-$R$ limit 
\cite{hyperuniformity2,hyperuniformity3}.
One should note that, 
in a monodisperse system of hard (non)spherical particles, 
the behavior of ${\cal S}(k)$ and 
that of ${\hat{\chi}}(k)$ 
in the limit $k \rightarrow 0$ 
are interconnected:
the two types of hyperuniformity, 
that of a point pattern and that of a two-phase medium, 
are either both absent or both present. 

One additional important quantity that characterizes a two-phase medium is
the pore-size distribution function ${\mathcal{P}}(\delta)$ 
together with 
its first, $\langle \delta \rangle$, 
and 
second, $\langle \delta^2 \rangle$, moments.
Here, 
$\delta$ is the radius of a hard sphere,       
completely and randomly insertable into 
the ${\mathcal{V}}_1$ phase.

The quantities $S_2(x)$, $\langle \delta \rangle$, and $\langle \delta^2 \rangle$ can lead to 
an estimate of the effective electromagnetic, mechanical, and transport properties of 
a random heterogeneous material made of phases ${{\cal V}_1}$ and ${\cal V}_2$
\cite{bookhetmat}.

\section{lens packing generation}
\label{packgen}

Due to the appreciably reduced propensity of optimal lenses to 
positionally and orientationally order \cite{lens2},
packings of $N=1013$ of them,
each with a surface area $S=2\sigma^2$,
with $\sigma$ the unit of length,
were progressively generated by 
gently compressing
the low-density equilibrium isotropic fluid phase until reaching 
the high-density nonequilibrium MRJ state. 
This compression was carried out via
an isobaric(-isothermal) Monte Carlo method-based procedure using
a triclinic computational box of volume $V$ and variable shape and size, and
periodic boundary conditions \cite{smallcomment}.
This allows one to closely monitor how
the (micro)structure of these packings changes in 
the process of formation of the MRJ state.

\section{results}
\label{results}

\begin{figure}
\psfrag{inverse compressibility factor}{\Large{$\displaystyle \frac{\varrho}{\beta P}$}}
\psfrag{packing fraction}{\Large{$\phi$}}
\psfrag{phm}{$\phi_{\rm MRJ}$}
\psfrag{phd}{$\phi_{\rm DKP}$}
\includegraphics[width=0.5\textwidth]{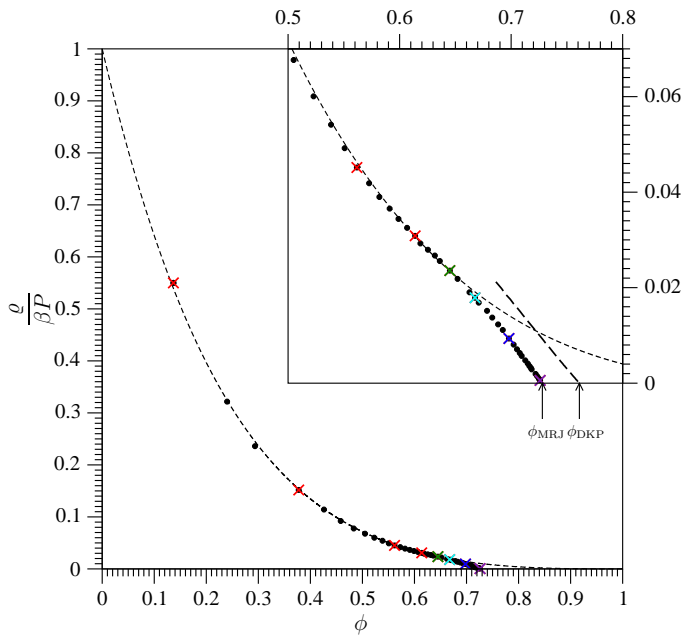}
\caption{
Inverse compressibility factor $\displaystyle \frac{\varrho}{\beta P}$ 
as a function of 
packing fraction $\phi$ (black circles).
Those marked with a colored 
cross are 
the state points at which 
a detailed analysis of the (micro)structure was carried out.
These state points are representative of 
the equilibrium fluid (red), metastable fluid (green), glassy (cyan and blue) and MRJ (indigo) states. 
The short-dashed curve is the analytic equation of state of Eq. \ref{eos81}.
The inset focuses on 
the high-$\phi$ regime where 
the high-$\phi$ equation of state for the crystal phase is also included 
as a long-dashed curve and
the two vertical arrows indicate the values of $\phi$ for the 
maximally random jammed (MRJ) state and 
the densest-known packings (DKP).
}
\label{figura2}
\end{figure}

One very important attribute of any hard-particle packing is 
its packing fraction $\phi=\varrho v = N/V v $, 
where $v$ is the particle volume.
The change in $\phi$ as 
a dilute equilibrium isotropic fluid system of optimal lenses is
gently compressed until 
reaching the nonequilibrium MRJ state 
is shown in Fig. \ref{figura2}.
In this figure,
the inverse compressibility factor, $\displaystyle \frac{\varrho}{\beta P}$, is plotted as 
a function of $\phi$. 
The monotonic gentle descent of 
$\displaystyle \frac{\varrho}{\beta P}$ bends further downwards at $\phi \approx 0.65$. 
Then, it continues essentially linearly, 
in accordance to free-volume theory (fvt) \cite{freevolumetheory},
until the MRJ state is reached at 
the fvt-extrapolated value $\phi_{\rm MRJ} \simeq 0.73$. 
This bend is particularly well appreciated by comparing 
the numerical simulation data to 
a past analytic equation of state proposed for 
the isotropic fluid phase of monodisperse systems of hard convex nonspherical particles
\cite{eosczech}:
\begin{equation}
\frac{\beta P}{\varrho} = \frac{1}{1-\phi} + \frac{3\alpha\phi}{(1-\phi)^2} + 
\frac{3\alpha^2\phi^2 -\alpha(6\alpha -5)\phi^3}{(1-\phi)^3}
\label{eos81}
\end{equation}
where $\displaystyle \alpha=\bar{R}S/(3v)$ is 
a nonsphericity parameter written in terms of 
the mean curvature radius $\bar{R}$, $S$ and $v$.
This analytic equation of state works very well within
the equilibrium fluid and the metastable fluid states but, 
doomed 
by the unphysical pole at $\phi=1$,
significantly departs from the numerical simulation data in 
the glassy and MRJ states.
The value of $\phi_{\rm MRJ} \simeq 0.73$ is only 4\% smaller than
the value of $\phi_{\rm DKP} = 0.76210\cdots$ \cite{lens1}.
For a given dimensionality $d$ of the Euclidean space,
the closer the value of the ratio $\displaystyle \phi_{\rm MRJ}/\phi_{\rm DKP}$ 
is to unity
the greater the propensity of 
a monodisperse system of hard particles is
to form mechanically stable glassy states.
The equilibrium crystal phase equation of state \cite{lens1} starts 
from the value of $\phi_{\rm DKP}$. 
The corresponding $\displaystyle \frac{\varrho}{\beta P}$ behaves 
essentially linearly as a function of $\phi$,
in accordance to fvt \cite{freevolumetheory}.
Its slope is similar to that of 
$\displaystyle \frac{\varrho}{\beta P}$ versus $\phi$ in 
the nonequilibrium glassy state.

In the following subsections, 
the (micro)structure of 
packings  
representative of 
the equilibrium fluid, metastable fluid, glassy and MRJ states are characterized via 
the structural descriptors from Section \ref{packchar}.


\subsection{Real-space pair correlation functions and reciprocal space structure factor}
\begin{figure}
\psfrag{r/s}{\large{$r/\sigma$}}
\psfrag{a/s}{\large{$a/\sigma$}}
\psfrag{b/s}{\large{$b/\sigma$}}
\psfrag{d/s}{\large{$\bar{\rm d}/\sigma$}}
\psfrag{gdr}{\large{$g(r)$}}
\psfrag{pdi}{\large{${\rm p}({\rm d})$}}
\psfrag{di/s}{\large{${\rm d}/\sigma$}}
\psfrag{phi=0.614}{\large{$\phi=0.614$}}
\psfrag{phi=0.645}{\large{$\phi=0.645$}}
\psfrag{phi=0.668}{\large{$\phi=0.668$}}
\psfrag{phi=0.698}{\large{$\phi=0.698$}}
\psfrag{phi=0.728}{\large{$\phi=0.728$}}
\includegraphics[width=0.5\textwidth]{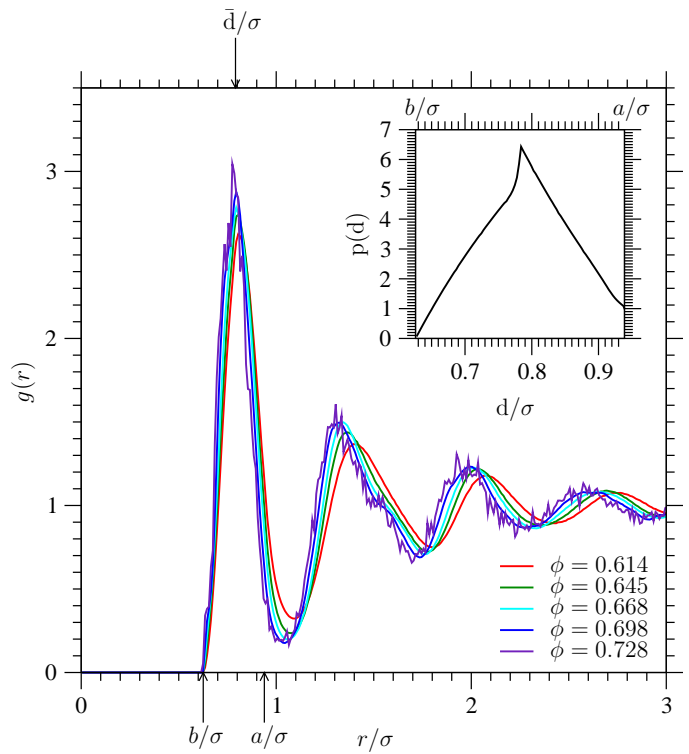}
\caption{
Pair correlation function $g(r)$ at 
several values of packing fraction $\phi$ representative of 
the equilibrium fluid (red), 
metastable fluid (green), 
glassy (cyan and blue) and  
MRJ (indigo) states.
The arrows indicate the value of: 
$a$, one of the optimal-lens $\mathsf{C}_2$ axes; 
$b$, the optimal-lens $\mathsf{C}_{\infty}$ axis; 
$\bar{\rm d}$, the \textit{in vacuo} mean contact distance.
The inset shows 
the optimal-lens \textit{in vacuo} probability density function of the contact distance
${\rm p} ({\rm d})$ 
i.e. 
the probability density to find two randomly chosen optimal lenses whose 
contact distance $\in [{\rm d}, {\rm d}+d{\rm d}]$.
}
\label{legr}
\end{figure}
The positional pair correlation function $g(r)$ is
the most basic function that describes the (micro)structure of 
a statistically homogeneous and isotropic system
\cite{liquid,hardspheres1,hardspheres2,hardspheres3,hardspheres4,hardspheres2bis,hardspheres5,bookhetmat}.
This function is given in Fig. \ref{legr} at 
several values of $\phi$ 
from the dense equilibrium fluid phase to the nonequilibrium MRJ state.
These $g(r)$'s have the form that this function typically takes on in 
the dense fluid state of a system composed of
hard moderately nonspherical particles.
The positional disordered character of the system is revealed, globally, by
the fast damped-exponential peak decay and valley rise towards 
the long-distance limit value of unity.
In addition,  
the principal peak abscissa value essentially remains stuck at 
$\simeq \bar{\rm d} = \int_b^a {\rm d p(d)} d{\rm d} = 0.791\dots \sigma$,  
the value of the \emph{in vacuo} mean contact distance 
(main and inset panels of Fig. \ref{legr}).
This fact suggests that
these hard particles, even locally, do not generally have a preferred organisation.
The principal peak abscissa value is moderately, yet perceptibly, moving towards 
the value of the $\emph{in vacuo}$ most probable contact distance, $0.784\dots \sigma$ (inset of Fig. \ref{legr}), 
as $\phi$ increases, which causes more contacts to be established between the hard particles.
The successive peaks shift towards the principal peak while
the principal valley deepens as $\phi$ increases.
During this compression,
there is no evident sign of the system becoming glassy except,
in retrospect,
the moderate displacement of the principal peak abscissa value,
the appearance of a tenuous shoulder at $r \approx 1.5 \sigma$
and, especially, the progressive roughness of the curve.
This roughness is a direct consequence of the rigidity that 
the system is acquiring and that 
its relatively small size makes noticeable \cite{notagrsk}.
The form of $g(r)$ for a system of optimal lenses in a disordered state differs from 
that of 
a system of hard spheres in a disordered state.
They differ
not just in the principal peak shape 
but especially in their overall smoothness as the MRJ state is approached.
Due to the nonsphericity of the hard particles, 
the principal peak is rounded off rather than spiky.
It is reminiscent 
of $g(r)$ of
a monodisperse system of soft (attractive-repulsive) spherical (e.g. Lennard-Jones) particles in
its liquid phase \cite{liquid,hardspheres1}.
Due again to the nonsphericity of the hard particles,
the form of this function as the MRJ state is approached 
does not show any singularities 
nor a split second peak,
both features of the hard-sphere MRJ state $g(r)$ 
\cite{hardspheres4,hardspheres5}.

Directly connected to $g(r)$ is 
the (orientationally averaged) structure factor ${\mathcal{S}}(k)$.
\begin{figure}
\psfrag{S(k)}{${\mathcal{S}}(k)$}
\psfrag{ks}{$k\sigma$}
\psfrag{k2s2}{$k^2\sigma^2$}
\includegraphics{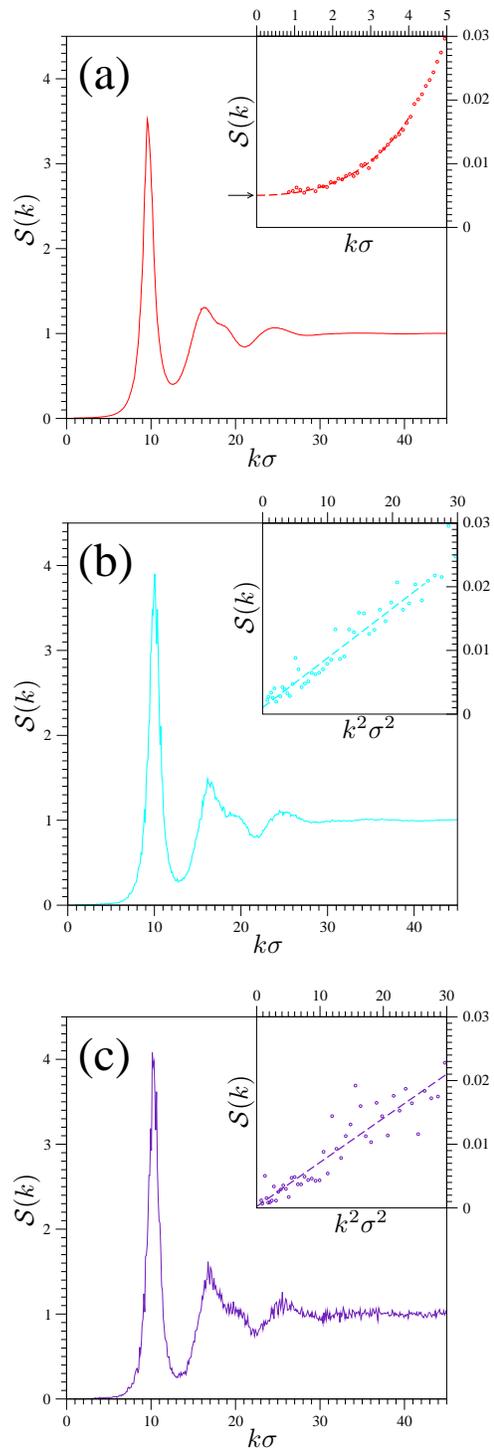}
\caption
{
Orientationally averaged 
structure factor ${\mathcal{S}}(k)$ at several values of packing fraction $\phi$:
(a) equilibrium fluid at $\phi$=0.614;
(b) glassy state at $\phi$=0.668;
(c) MRJ state at $\phi$=0.728.
In any panel, 
the right top inset focuses on the low-$k$ regime with the dashed curve being a quadratic fit.
In (a)
the arrow marks the value of ${\mathcal{S}(}0)$ obtained from 
the isothermal compressibility;
in (b) and (c) it had better plot ${\mathcal{S}}(k)$ as a function of $k^2$ so 
as to more clearly show its diminishing trend as $k \rightarrow 0$.
}
\label{struttofat}
\end{figure}
The overall form of ${\mathcal{S}}(k)$, 
particularly its limit value of unity as $k \rightarrow \infty$,
at values of $\phi$ in the dense equilibrium fluid, nonequilibrium glassy and MRJ states 
confirms  
the positionally disordered character of all these states.  
The strong similarity among all these curves indicates that 
these states are cognate with one another (Fig. \ref{struttofat}). 
In parallel to what observed for $g(r)$ (Fig. \ref{legr}), 
the progressive roughness of the curve as $\phi$ increases (Fig. \ref{struttofat}) is 
a reflection of 
the progressive rigidity that
the system is acquiring and that 
the small size of the system makes noticeable \cite{notagrsk}.
In the equilibrium fluid state, 
the value of ${\mathcal{S}}(0) > 0$ obtained by 
quadratically fitting the low-$k$ ${\mathcal{S}}(k)$ data 
matches the value obtained from the isothermal compressibility [Fig. \ref{struttofat} (a)],
which, in equilibrium, is known to be related to ${\mathcal{S}}(0)$ \cite{liquid}.
In the denser nonequilibrium states, 
the extrapolated value of ${\mathcal{S}}(0)$ keeps progressively decreasing [Fig. \ref{struttofat} (b,c)].
With the caveat that
the present system size is not so large to allow for 
very small values of $k$ to be investigated and
the statistics of ${\cal S}(k)$ at these very small $k$'s to be extremely good,
one may conclude that 
the values that ${\mathcal{S}}(k)$ takes on as $k \rightarrow 0$ are 
so small [Fig. \ref{struttofat} (c)] that
the system becomes effectively hyperuniform on approaching the MRJ state: 
in the neighborhood of $k = 0$
${\mathcal{S}} \simeq 10^{-3}$ while 
${\mathcal{S}} \simeq 4$
at its principal peak \cite{hyperuniformity3}.

The nonsphericity of the present hard particles offers
the possibility to define and evaluate
new, orientational, pair correlation functions
as well as more precise bond-orientational pair correlation functions.
Given the cylindrically symmetric character of the present hard particles, 
$\mathcal{G}_{2}^{\hat{\mathbf{u}}\hat{\mathbf{u}}} (r)$ and
$\mathcal{G}_{2}^{\hat{\mathbf{u}}\hat{\mathbf{r}}} (r)$ 
are 
the most basic orientational and bond-orientational pair correlation functions. 
\begin{figure*}
\psfrag{r/s}{\Large{$r/\sigma$}}
\psfrag{Guprp}{\Large{$\mathcal{G}_{2}^{\hat{\mathbf{u}}\hat{\mathbf{u}}} (r)$}}
\includegraphics{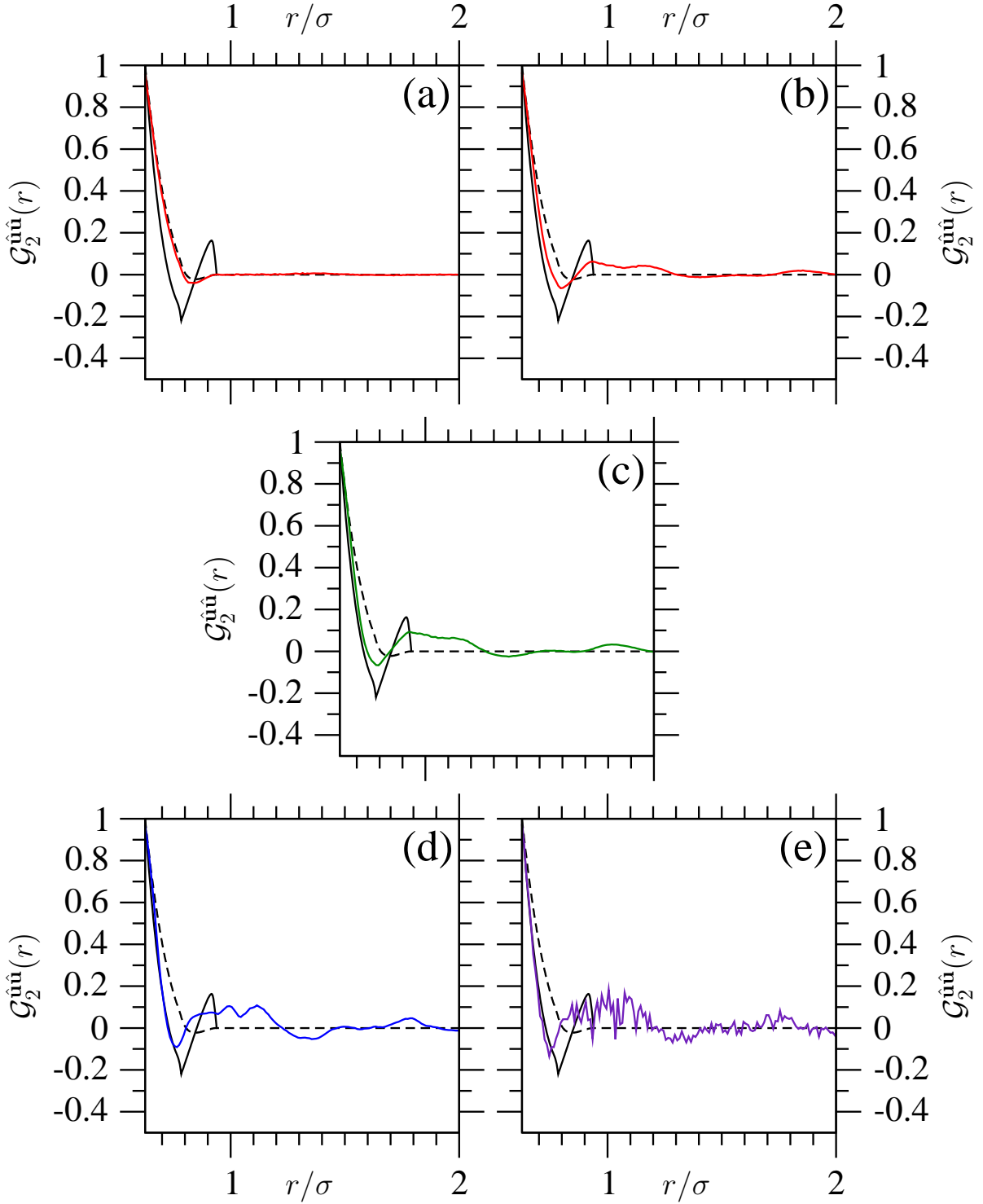}
\caption
{
Pair correlation function 
$\mathcal{G}_{2}^{\hat{\mathbf{u}}\hat{\mathbf{u}}} (r)$
at several values of packing fraction $\phi$ representative of 
the equilibrium fluid (a, $\phi=0.378$; b, $\phi=0.614$), 
metastable fluid (c, $\phi=0.645$), 
glassy (d, $\phi=0.698$) and MRJ (e, $\phi=0.728$) states.
In each panel, 
the black curves correspond to 
the limit form this function takes on 
when calculated considering two randomly chosen lenses
either free (dashed) or touching (continuous).
\label{g2uu}
}
\end{figure*}
\begin{figure*}
\psfrag{r/s}{\Large{$r/\sigma$}}
\psfrag{Grprp}{\Large{$\mathcal{G}_{2}^{\hat{\mathbf{u}}\hat{\mathbf{r}}} (r)$}}
\includegraphics{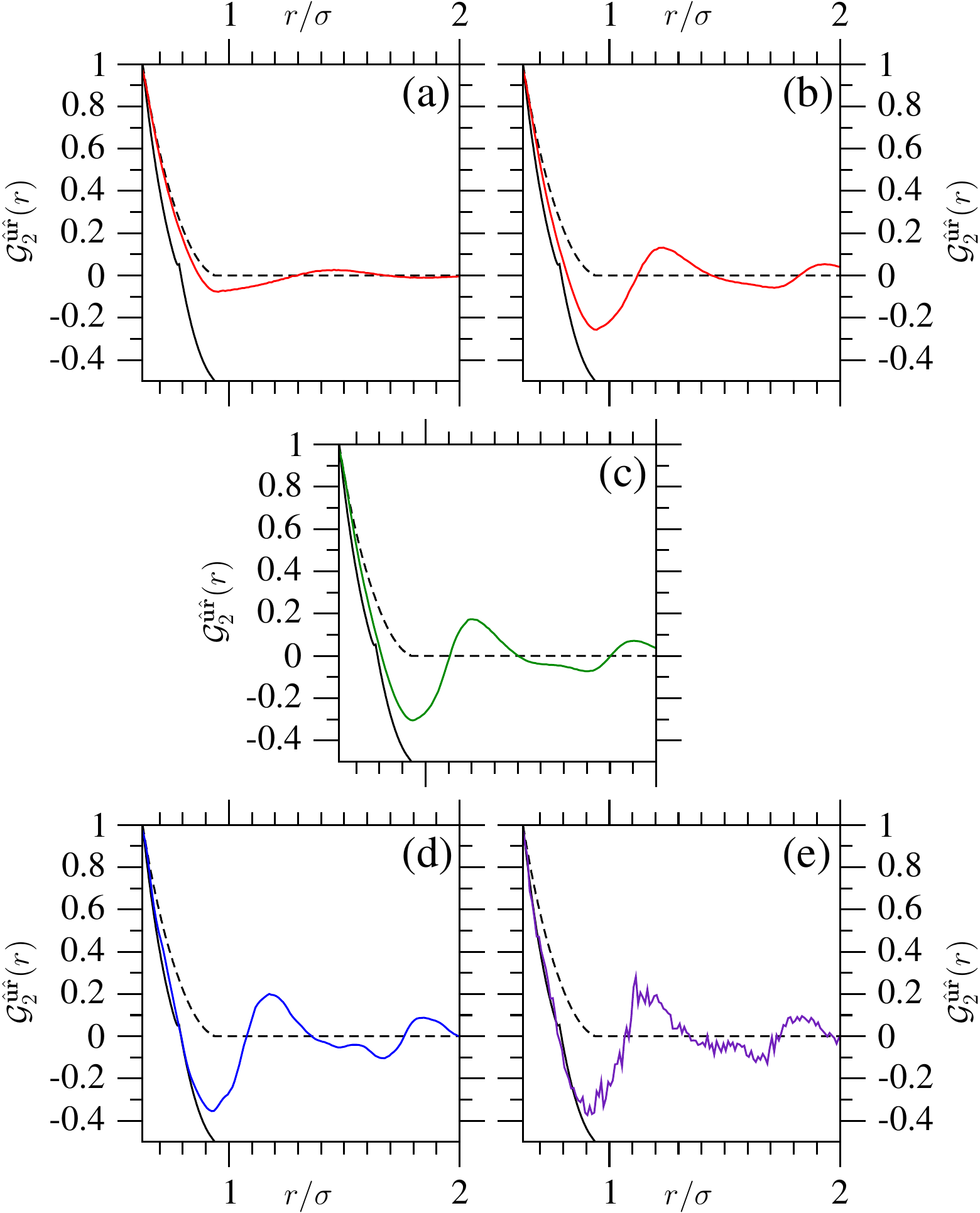}
\caption{
Pair correlation function 
$\mathcal{G}_{2}^{\hat{\mathbf{u}}\hat{\mathbf{r}}} (r)$
at several values of packing fraction $\phi$ representative of 
the equilibrium fluid (a, $\phi=0.378$; b, $\phi=0.614$), 
metastable fluid (c, $\phi=0.645$), 
glassy (d, $\phi=0.698$) and MRJ (e, $\phi=0.728$) states.
In each panel, 
the black curves correspond to 
the limit form this function takes on
when calculated considering two randomly chosen lenses
either free (dashed) or touching (continuous).
}
\label{g2ur}
\end{figure*}
Their form at several values of $\phi$,
from the moderately dense equilibrium fluid to the nonequilibrium MRJ states, 
are shown in Figs. \ref{g2uu} and \ref{g2ur}.
The vanishing of 
$\mathcal{G}_{2}^{\hat{\mathbf{u}}\hat{\mathbf{u}}} (r)$ and
$\mathcal{G}_{2}^{\hat{\mathbf{u}}\hat{\mathbf{r}}} (r)$ 
as $r\rightarrow \infty$ 
demonstates the globally orientationally disordered character of all considered packings.  
In each panel of these figures, 
$\mathcal{G}_{2}^{\hat{\mathbf{u}}\hat{\mathbf{u}}} (r)$ and
$\mathcal{G}_{2}^{\hat{\mathbf{u}}\hat{\mathbf{r}}} (r)$ 
are compared to 
two functions.
The first is the limit form that 
$\mathcal{G}_{2}^{\hat{\mathbf{u}}\hat{\mathbf{u}}} (r)$ and
$\mathcal{G}_{2}^{\hat{\mathbf{u}}\hat{\mathbf{r}}} (r)$ 
respectively take on as $\phi \rightarrow 0$. 
This corresponds to a calculation where 
two particles are taken at a fixed centroid distance $r$ and
whose orientations are completely random except that
the nonoverlap constraint has to be complied with.   
The domain of these functions is $[b,\infty)$.
The second is the limit form that 
$\mathcal{G}_{2}^{\hat{\mathbf{u}}\hat{\mathbf{u}}} (r)$ and
$\mathcal{G}_{2}^{\hat{\mathbf{u}}\hat{\mathbf{r}}} (r)$ 
respectively take on in a calculation where 
two particles are taken whose orientations are completely random
except that the particles are constrained to touch.
They are related to the form respectively taken on by  
$\mathcal{G}_{2}^{\hat{\mathbf{u}}\hat{\mathbf{u}}} (r)$ and 
$\mathcal{G}_{2}^{\hat{\mathbf{u}}\hat{\mathbf{r}}} (r)$ 
as the MRJ state is approached.
The domain of these functions is $[b,a]$.
One can observe that 
$\mathcal{G}_{2}^{\hat{\mathbf{u}}\hat{\mathbf{u}}} (r)$ and
$\mathcal{G}_{2}^{\hat{\mathbf{u}}\hat{\mathbf{r}}} (r)$ 
progressively pass from the respective first limit form and 
change so as 
to `adhere' to the respective second limit form
as $\phi$ increases.
Even in the dense packings, 
the second limit form cannot be completely `adhered' to
since for sufficiently large $r$ 
not all pairs of particles whose
centroids are separated by $r$ are necessarily touching. 
Nonetheless,
the second limit form sets a paragon stone
by which to understand how
$\mathcal{G}_{2}^{\hat{\mathbf{u}}\hat{\mathbf{u}}} (r)$ and
$\mathcal{G}_{2}^{\hat{\mathbf{u}}\hat{\mathbf{r}}} (r)$ 
changes as $\phi$ increases.
The fact that
$\mathcal{G}_{2}^{\hat{\mathbf{u}}\hat{\mathbf{u}}} (r)$ and
$\mathcal{G}_{2}^{\hat{\mathbf{u}}\hat{\mathbf{r}}} (r)$ 
are taking on 
a form that closely resembles the respective second limit form is 
an indication that the packings are also locally orientationally disordered.
Even when only viewing 
$\mathcal{G}_{2}^{\hat{\mathbf{u}}\hat{\mathbf{u}}} (r)$ and
$\mathcal{G}_{2}^{\hat{\mathbf{u}}\hat{\mathbf{r}}} (r)$,
the process of formation of the nonequilibrium MRJ state from the equilibrium fluid state
is one in which 
the salient features of the (micro)structure `exasperate' quantitatively,
as the degree of contactedness between the particles a fortiori increases,
without however significantly changing qualitatively. 

\subsection{Pair correlation function of the scaled distance and contact statistics}
\begin{figure}
\psfrag{phi=0.137}{\large{$\phi=0.137$}}
\psfrag{phi=0.378}{\large{$\phi=0.378$}}
\psfrag{phi=0.614}{\large{$\phi=0.614$}}
\psfrag{phi=0.645}{\large{$\phi=0.645$}}
\psfrag{phi=0.668}{\large{$\phi=0.668$}}
\psfrag{phi=0.698}{\large{$\phi=0.698$}}
\psfrag{phi=0.728}{\large{$\phi=0.728$}}
\psfrag{gds}{\large{${\mathtt{g}}(s)$}}
\psfrag{sss}{\large{ $s$} }
\includegraphics[width=0.5\textwidth]{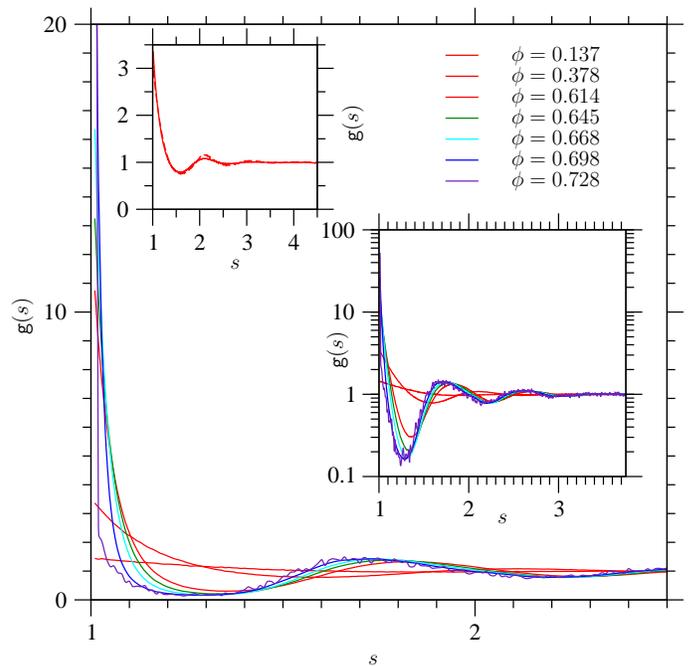}
\caption
{
Pair correlation function ${\mathtt{g}}(s)$ of the scaled distance $s$,
obtained by dividing the distance $r$ separating two lens centroids by the appropriate contact distance,
at several values of packing fraction $\phi$ representative of
the equilibrium fluid (red), metastable  (green), glassy (cyan and blue) and MRJ (indigo) states.
The top left inset shows ${\mathtt{g}}(s)$ in 
the moderately dense equilibrium fluid phase at $\phi=0.378$ (continuous curve) compared to 
the Percus-Yevick approximation result for 
the $g(r)$ of the hard-sphere fluid at the same value of $\phi$ (dashed curve).
The bottom right inset shows the same ${\mathtt{g}}(s)$'s as the main panel but with 
the ordinate axis on a logarithmic scale.
}
\label{lags}
\end{figure}

The moderate nonsphericity of the present hard particles is responsible for 
$g(r)$ having 
a form resembling more that of liquid argon \cite{liquid,hardspheres1} rather than 
that of 
the hard-sphere fluid 
\cite{liquid,hardspheres1,hardspheres2,hardspheres3,hardspheres4,hardspheres2bis,hardspheres5,bookhetmat}.
Considering the scaled distance 
$s=r/{\cal D}\left({\hat{\mathbf{r}}}_{ij}, {\hat{\mathbf{u}}}_i, {\hat{\mathbf{u}}}_j \right)$
instead of 
the real distance $r$ 
restores a pair correlation function ${\mathtt{g}}(s)$ with 
a ``hard-sphere-fluid--like'' form (Fig. \ref{lags}).
By construction, 
as $\phi \rightarrow 0$,  
${\mathtt{g}}(s)$ is guaranteed 
to approach 
the corresponding unit-diameter hard-sphere fluid $g(r)$, $g_{\rm hs}(r)$,
i.e. the step function
\begin{eqnarray}
\mathsf{\Theta}(r)=\left\{ 
\begin{array}{lr}
0  : &   
0 \le r \le 1 \\
1  : & 
r  > 1
\end{array}
\right.
.
\end{eqnarray}
It becomes of interest to investigate how 
${\mathtt{g}}(s)$ compares to $g_{\rm hs}(r)$ as $\phi$ increases.
This may be done by using, for the hard-sphere positional pair correlation function,
the Percus-Yevick (PY) approximation, $g_{\rm hs}^{\rm PY} (r)$,
known to be good throughout 
the hard-sphere equilibrium fluid phase \cite{liquid,hardspheres1,hardspheres2,hardspheres2bis}.
Indeed, $g_{\rm hs}^{\rm PY} (r)$ compares well to ${\mathtt{g}}(s)$ up to 
moderate values of $\phi$ (top-left inset of Fig. \ref{lags}).
However, the two positional pair correlation functions progressively depart from one another as 
$\phi$ increases and surpasses $\phi_{\rm hs, frz} = 0.494$,
the value of $\phi$ at which the hard-sphere fluid freezes.
Beyond $\phi_{\rm hs, frz}$, the PY approximation quickly deteriorates to such an extent that,
in the proximity of $\phi_{\rm hs, MRJ} \simeq 0.64$,
the value of $\phi$ at the hard-sphere MRJ state \cite{mrj},
$g_{\rm hs}^{\rm PY} (r)$ displays unphysically negative values.
The `true' $g_{\rm hs}(r)$ progressively loses a fluid-like appearance to 
finally assume the characteristic form with a singular split second peak that 
it exhibits at the MRJ state \cite{hardspheres3,hardspheres4,hardspheres5}.
On the contrary, 
${\mathtt{g}}(s)$ smoothly changes as $\phi$ increases towards $\phi_{\rm MRJ}$,
always maintaining a `hard-sphere-fluid--like' form.
Indeed, ${\mathtt{g}}(s)$ is tending to acquire 
an approximate `delta-plus-step-with-a-gap' form rather than 
the form characteristic of the three-dimensional hard-sphere MRJ state.
This is an example of 
the decorrelation principle that is acting as $d_{\rm f}$ increases
either because the dimensionality $d$ of the Euclidean space increases \cite{highdimhs} and/or
rotational degrees of freedom are added.
One can compare the abscissa value of
the minimum of ${\mathtt{g}}(s)$ at the MRJ state, $s\approx 1.3$, 
with
the optimal value of the gap parameter $\sigma^*$ discussed in 
the analysis of three- and higher-dimensional disordered hard-sphere systems aimed at
estimating the scaling of $\phi_{\rm MRJ}$ as $d \rightarrow \infty$ \cite{highdimhs}.

\begin{figure*}
\includegraphics{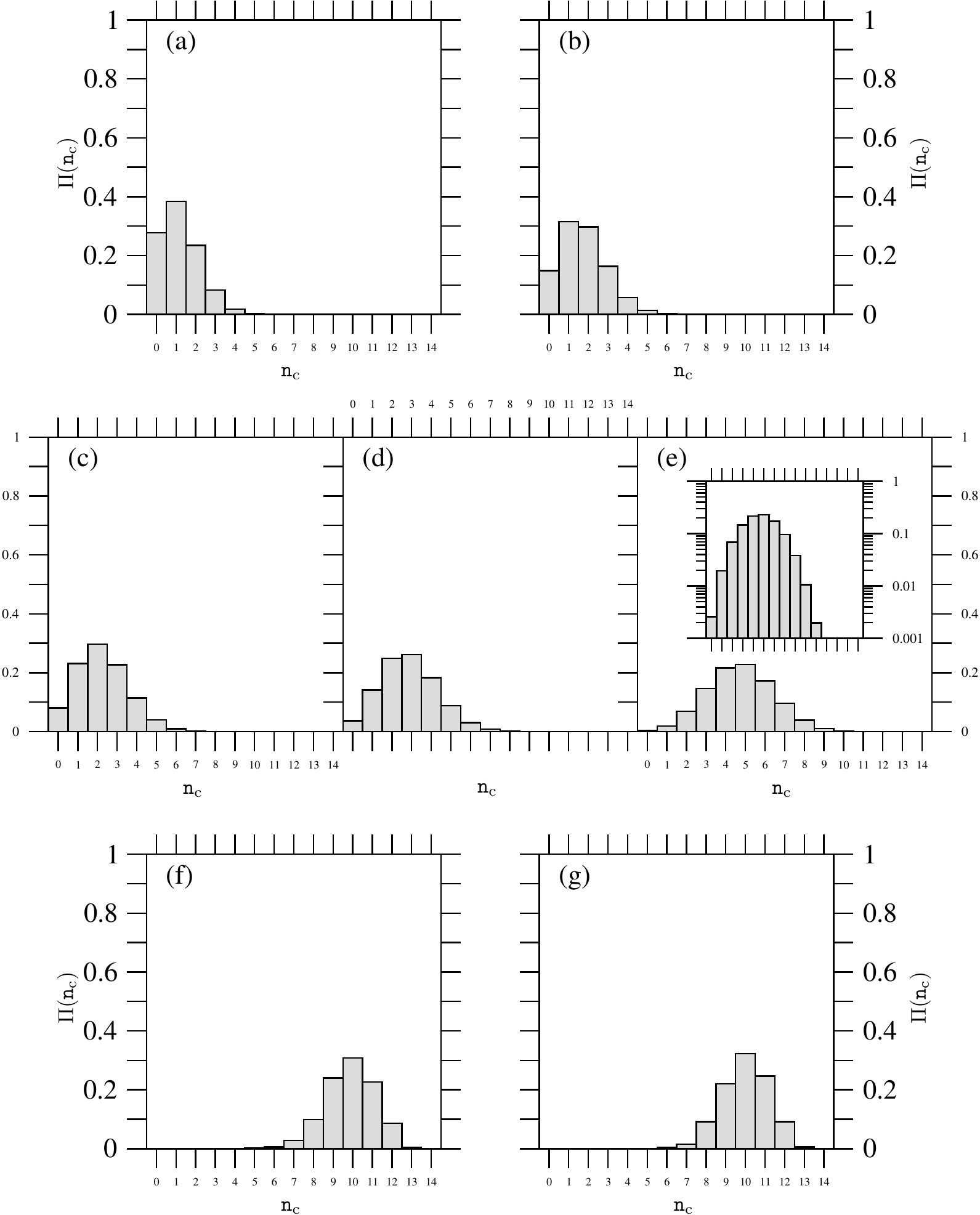}
\caption{
contact neighbor number probability $\Pi({\mathtt{n}}_{\rm c})$ at several values of packing fraction $\phi$
representative of the equilibrium fluid [(a), $\phi=0.562$; (b), $\phi=0.614$], 
metastable fluid [(c), $\phi=0.645$], 
glassy [(d), $\phi=0.668$; (e), $\phi=0.698$] 
and (quasi)MRJ states [(f), $\phi=0.726$; (g), $\phi=0.728$]. 
In (e), the inset reproduces the main panel with the ordinates on a logarithm scale.
}
\label{Pidinc}
\end{figure*}

\begin{figure}
\psfrag{Zeta}{\Large{$\mathtt{Z}$}}
\psfrag{volume fraction}{\Large{$\phi$}}
\includegraphics[width=0.5\textwidth]{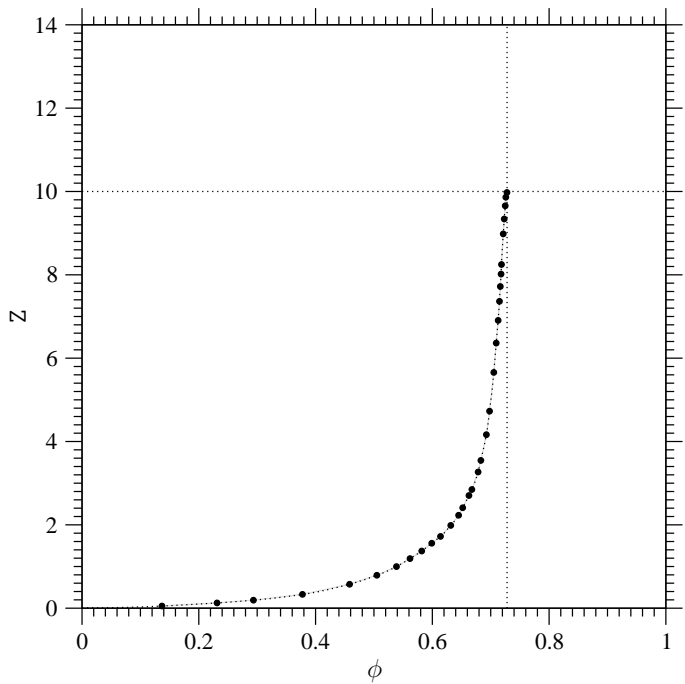}
\caption{
mean contact neighbor number $\mathtt{Z}$ as a function of packing fraction $\phi$ 
(black circles and guide-to-the-eye dotted curve).
The horizontal and vertical dotted lines intercept at ($\phi_{\rm MRJ}$,10).
}
\label{lozeta}
\end{figure}

The numerical calculation of ${\mathtt{g}}(s)$ leads to 
the calculation of 
the number of neighboring particles that are in contact with a central particle.
In fact, this number $\mathtt{n}_{\rm c}$ is defined here as 
the number of particles whose $s \in [1, 1+ds]$ with $ds = 0.01$. 
By averaging over central particles and configurations, 
one can calculate the probability, $\Pi(\mathtt{n}_{\rm c})$, that 
a particle has $\mathtt{n}_{\rm c}$ contact neighbors.
The histograms of $\Pi(\mathtt{n}_{\rm c})$ at several values of $\phi$
from the dense equilibrium fluid to the nonequilibrium MRJ states are shown in Fig. \ref{Pidinc}. 
During this compression, 
besides the expected progressive increase of 
the mean value of $\mathtt{n}_{\rm c}$, 
$\left \langle \mathtt{n}_{\rm c} \right \rangle = \mathtt{Z} $ (Fig. \ref{lozeta}), 
and that of
the most probable value of $\mathtt{n}_{\rm c}$,
the form of the histograms passes from being left- to right-skewed.
This fact decisively contributes to the upswing of $\mathtt{Z}$ in the proximity of the MRJ state until
it reaches the isostatic mean value of 10 at the MRJ state (Fig. \ref{lozeta}).
During this compression, the number of `rattlers', 
i.e. the particles with no contact neighbors, $\mathtt{n}_{\rm c} = 0$, 
quickly diminishes and vanishes in the close proximity of the MRJ state.
This occurs at 
the setting in of fully glassy behavior, 
in turn corresponding to 
the setting in of a fvt-like linear behavior of $\varrho/\beta P$ versus $\phi$ (Fig. \ref{figura2}).
The number of `rattling' optimal lenses is rather large at 
values of $\phi \simeq \phi_{\rm hs, MRJ} \simeq 0.64$ \cite{mrj}.
This is consistent with the capability of a system of optimal lenses  
to form 
an equilibrium fluid denser than 
the densest hard-sphere equilibrium fluid at $\phi_{\rm hs, frz}$.
It is also consistent with the capability of a system of optimal lenses  
to reach 
a MRJ state
not only $\simeq 14\%$ denser than the hard-sphere MRJ state 
but also remarkably devoid of any `rattler'.
Thus far, no procedure has been able to generate a `rattler'-free three-dimensional hard-sphere MRJ state
\cite{hardspheres4,hardspheres5,hyperuniformity3}.

\subsection{Lens packings as two-phase media}
\subsubsection{Two-point correlation function and spectral density}
\begin{figure}
\psfrag{phi=0.137}{\large{$\phi=0.137$}}
\psfrag{phi=0.378}{\large{$\phi=0.378$}}
\psfrag{phi=0.614}{\large{$\phi=0.614$}}
\psfrag{phi=0.645}{\large{$\phi=0.645$}}
\psfrag{phi=0.668}{\large{$\phi=0.668$}}
\psfrag{phi=0.698}{\large{$\phi=0.698$}}
\psfrag{phi=0.728}{\large{$\phi=0.728$}}
\psfrag{S2extdr}{\large{$\Sigma_{2_{\rm ext}}(x)$}}
\psfrag{r/s}{\large{$x/\sigma$}}
\includegraphics[width=0.5\textwidth]{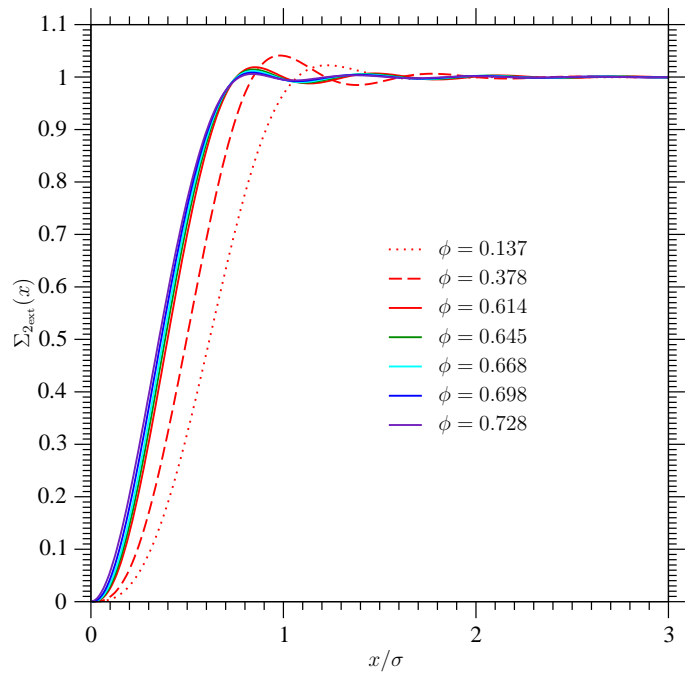}
\caption{
The external part of 
the two-point correlation function $\Sigma_{2{\rm ext}}(x)$ at 
several values of packing fraction $\phi$
representative of 
the equilibrium fluid (red; dotted, dashed and continous),
metastable fluid (green), 
glassy (cyan and blue) and 
MRJ (indigo) states.
}
\label{Sigma2ext}
\end{figure}
\begin{figure}
\psfrag{phi=0.137}{\large{$\phi=0.137$}}
\psfrag{phi=0.378}{\large{$\phi=0.378$}}
\psfrag{phi=0.614}{\large{$\phi=0.614$}}
\psfrag{phi=0.645}{\large{$\phi=0.645$}}
\psfrag{phi=0.668}{\large{$\phi=0.668$}}
\psfrag{phi=0.698}{\large{$\phi=0.698$}}
\psfrag{phi=0.728}{\large{$\phi=0.728$}}
\psfrag{ks}{\large{$k\sigma$}}
\psfrag{Xdk/s3}{\large{${\hat{\chi}}(k) \sigma^{-3}$}}
\includegraphics[width=0.5\textwidth]{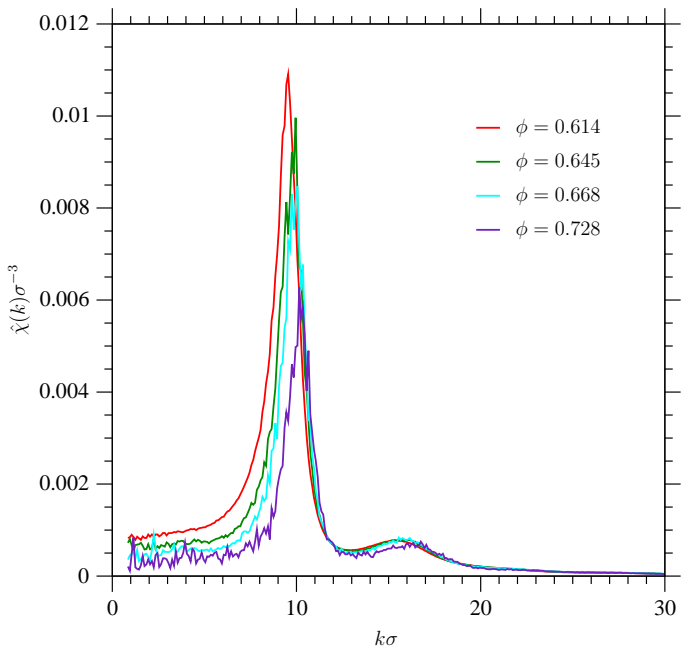}
\caption{
Spectral density ${\hat{\chi}}(k)$ at 
several values of packing fraction $\phi$ 
representative of
the equilibrium fluid (red), metastable fluid (green),
glassy (cyan) and MRJ (indigo) states.
}
\label{spectra}
\end{figure}

 
The most important component of the two-point probability function $S_2(x)$ is 
the external pair correlation function $\Sigma_{2{\rm ext}}(x)$.
This function is proportional to 
the conditional probability to find the two points at 
a distance $x$ and inside two different particles (Fig. \ref{Sigma2ext}).
Irrespective of the value of $\phi$, 
this function rather quickly reaches its $x \rightarrow \infty$ limit value of unity.
On increasing $\phi$, 
$\Sigma_{2{\rm ext}}(x)$ is expectedly progressively `pushed' towards $x=0$ while
growing damped oscillations are developed.
They show the largest amplitude in the moderately dense equilibrium fluid phase.
Then, these damped oscillations progressively fade away 
as $\phi$ approaches the value corresponding to the equilibrium fluid phase at freezing. 
From this point,
on further increasing $\phi$, 
$\Sigma_{2{\rm ext}}(x)$ moderately changes its form: 
it keeps being `pushed' mildly towards $x=0$ and reduces its damped oscillations.
This is diametrically opposed to 
what happens to $g(r)$ whose 
damped oscillations increase with $\phi$.
This suggests that 
an analytic theory that reliably extrapolates
$\Sigma_{2{\rm ext}}(x)$ to $x \rightarrow \infty$ might be more feasible than 
an analogous analytic theory for $g(r)$. 
That analytic theory would allow one to calculate ${\hat{\chi}}(k)$ by 
Fourier transform even for $k \rightarrow 0$. 
Short of such an analytic theory, 
${\hat{\chi}}(k)$ has directly been calculated (Fig. \ref{spectra}).
Leaving aside the expected progressive lowering of the curve as $\phi$ increases,
the overall form of ${\hat{\chi}} (k)$ changes little as 
the system goes from the equilibrium fluid to the nonequilibrium MRJ states.

\subsubsection{Pore-size statistics}
\begin{figure}
\psfrag{phi=0.137}{\large{$\phi=0.137$}}
\psfrag{phi=0.378}{\large{$\phi=0.378$}}
\psfrag{phi=0.614}{\large{$\phi=0.614$}}
\psfrag{phi=0.645}{\large{$\phi=0.645$}}
\psfrag{phi=0.668}{\large{$\phi=0.668$}}
\psfrag{phi=0.698}{\large{$\phi=0.698$}}
\psfrag{phi=0.728}{\large{$\phi=0.728$}}
\psfrag{del}{\large{$\delta/\sigma$}}
\psfrag{Pde}{\large{${\mathcal{P}}(\delta) \sigma$}}
\includegraphics[width=0.5\textwidth]{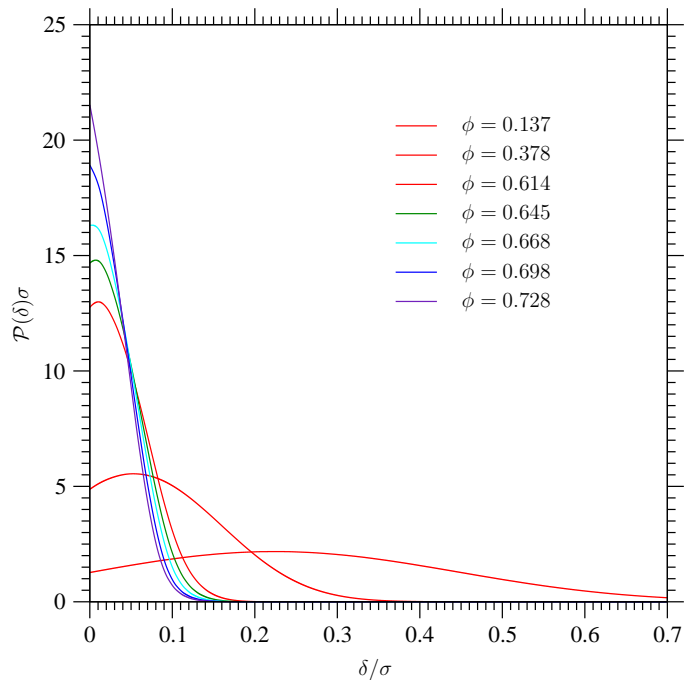}
\caption{
Pore-size distribution function ${\mathcal{P}}(\delta)$ at several value of packing fraction $\phi$,
from the low-$\phi$ equilibrium fluid to the high-$\phi$ MRJ state.
Red curves correspond to the equilibrium fluid state;
the green curve corresponds to the nonequilibrium fluid state;
the cyan and blue curves correspond to the glassy state;
the indigo corresponds to the MRJ state.
\label{poro}
}
\end{figure}

\begin{figure}
\psfrag{phi}{\Large{$\phi$}}
\psfrag{1del}{\Large{$\frac{\sigma}{\left \langle \delta \right \rangle}$}}
\psfrag{1del2}{\Large{$\frac{\sigma^2}{\left \langle \delta ^2 \right \rangle}$}}
\includegraphics{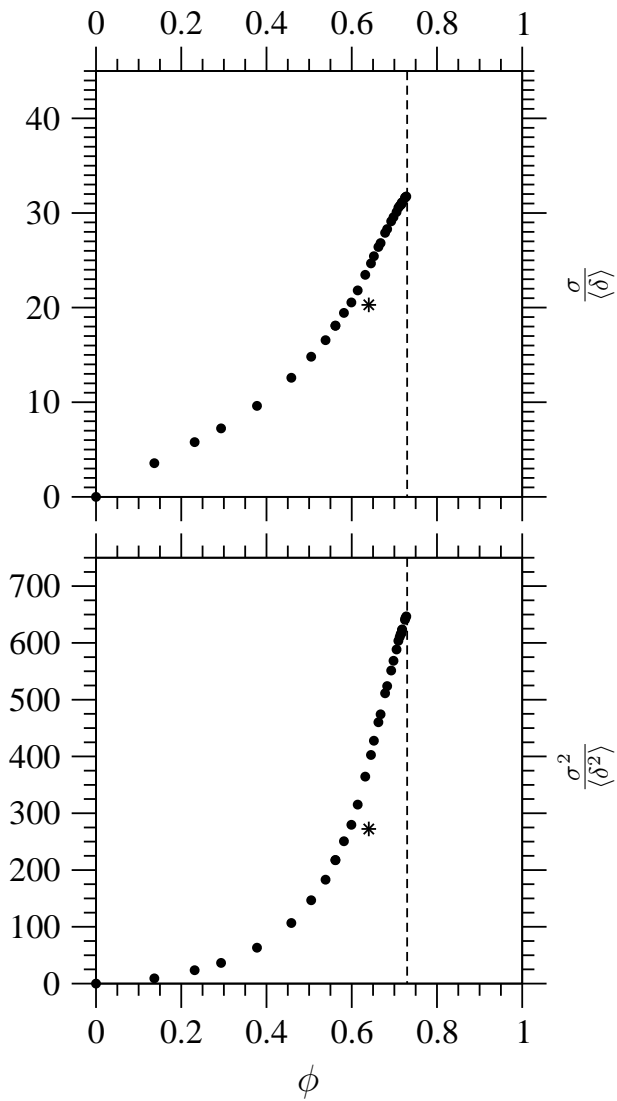}
\caption{
Inverse of mean pore size $\left \langle \delta \right \rangle$ and
mean square pore size $\left \langle \delta^2 \right \rangle$ as 
a function of packing fraction $\phi$.
The vertical dashed lines mark the the value of $\phi$ of the optimal-lens MRJ state.
The asterisks are the corresponding data for the hard-sphere MRJ state.
}
\label{momenti}
\end{figure}

One additional important quantity when characterizing a two-phase medium is 
its pore-size distribution function ${\mathcal{P}}(\delta)$ (Fig. \ref{poro}). 
Its form significantly changes when
going from the dilute equilibrium fluid to the dense nonequilibrium MRJ states. 
In addition to 
the expected progressive sharpening and shifting of the function $P(\delta)$  upward as 
$\delta$ tends to zero as $\phi$ increases, 
$P(\delta=0)$ is its maximum value at sufficiently high $\phi$
and (consequently) 
its derivative at $\delta=0$ changes from positive to negative.
This occurs 
in correspondence to 
the system becoming glassy and then reaching the MRJ state.
Either directly or from ${\mathcal{P}}(\delta)$, 
one can calculate the first two moments of this distribution function, 
$\left \langle \delta \right \rangle$ and $\left \langle \delta^2 \right \rangle$ (Fig. \ref{momenti}).
It proved important to report the inverse of these two quantities versus $\phi$ so as 
to reveal the quasisigmoidal form that these two curves have on approaching the MRJ state. 
This allows one to appreciate how 
the graph of $\phi$ versus $1/\left \langle \delta \right \rangle$ and 
that of $\phi$ versus $1/\left \langle \delta^2 \right \rangle$ 
mirror the graph of $\varrho/\beta P$ versus $\phi$ (Fig. \ref{figura2}):
the bend that these former curves have at $\phi \approx 0.65$ parallels
the bend that this latter curve has at the same value of $\phi$.
For the hard-sphere MRJ state at $\phi_{\rm hs, MRJ}\simeq 0.64$,
$\left \langle \delta \right \rangle \simeq 0.063 {\rm D}$ and 
$\left \langle \delta^2 \right \rangle \simeq 0.006 {\rm D}$,
with D the hard-sphere diameter \cite{hscharII}.
The latter length is presently assimilable to $\bar{\rm d}$.
Thus, for the hard-sphere MRJ state at $\phi_{\rm hs, MRJ}\simeq 0.64$,
$\left \langle \delta \right \rangle \simeq 0.049 \sigma$ and
$\left \langle \delta^2 \right \rangle \simeq 0.0037 \sigma^2$.
These values are significantly larger than
the respective value for optimal lenses at the same value of $\phi \simeq 0.64$:
another confirmation that optimal lenses are  
better (ordered and disordered) packers than hard spheres.

\section{conclusions}
\label{conclu}

In the class of hard convex lens-shaped particles, 
the member with aspect ratio equal to 2/3 is `optimal' in the sense that
its maximally random jammed state is the densest, 
which imparts them with a reduced propensity 
to positionally and/or orientationally order 
on compressing from the equilibrium isotropic fluid.
This makes them 
a suitable hard nonspherical particle model  
to carefully investigate the process of formation of the maximally random jammed state
without interference from not only full but also partial, plastic- or liquid-, crystallization while
keeping the system monodisperse.
Thus, by using a simple Monte Carlo method-based procedure,
monodisperse packings of such hard nonspherical particles are generated by compressing 
the dilute isotropic fluid until 
reaching the maximally random jammed  state.

To characterize how the (micro)structure of these packings changes in this process, 
many structural descriptors are calculated.
These structural descriptors undergo 
gradual, quantitative but not qualitative, changes:
the compression `exasperates' features that 
are already present in the dense equilibrium isotropic fluid.
These changes can coherently and consistenly be traced back to 
the gradual increase of contacts between these hard particles on densification until 
the isostatic mean value of 10 contact neighbors per particle is reached at 
the effectively hyperuniform maximally random jammed state.
Even the bend in the inverse compressibility factor versus packing fraction curve,
a macroscopic signature of glass formation,  
can be traced back to 
the pore-size distribution function assuming its absolute maximum at a pore size equal to zero.

The analysis of contact statistics can be seen as 
part of the calculation of 
the pair correlation function of the scaled distance obtained by 
dividing the real distance by the orientation dependent distance of closest approach.
The form of this special pair correlation function compares well to 
the one of a hard-sphere fluid up to moderate values of packing fraction.
For values of the packing fraction approaching and surpassing 
the value at hard-sphere freezing,
the two pair correlation functions depart more and more from one another. 
The hard-sphere pair correlation function is known to acquire 
a form distinct from the one in the equilibrium fluid, 
with a singular split second peak, 
as the maximally random jammed  is approached and finally reached. 
Instead, the pair correlation function of the scaled distance always maintains
a fluid-like form that approximates a `delta-plus-step-with-a-gap' form as 
the maximally random jammed state is approached and finally reached.
This can be seen as an example of the decorrelation principle 
acting as the number of degrees of freedom increases.

Compared to the hard-sphere maximally random jammed state,
the maximally random jammed state of the present hard nonspherical particles is 
not only denser but also has 
a packing fraction only a few percent smaller than the packing fraction of the corresponding densest-known crystalline (degenerate) packings.
Based on the decorrelation principle,
it can be considered more disordered. 
In addition, it is rattler-free and less porous.
These characteristics make it 
a significantly better glassy material and 
the investigation of 
its effective electromagnetic, mechanical and trasport properties \cite{bookhetmat,hscharIII} opportune.
It is possible that 
other hard convex uniaxial particle models with 
an aspect ratio equal to 2/3, if oblate, or 3/2, if prolate, 
might also be found `optimal' in the same sense used for lenses and
that moderate biaxial variants of them might form 
disordered packings with further improved characteristics.

\acknowledgments
The authors are grateful to Charles Maher for his careful reading of the manuscript.
G.C. acknowledges the support of the Government of Spain
under grants no. FIS2013-47350-C5-1-R,  no. MDM-2014-0377 and no. FIS2017-86007-C3-1-P  while
S.T. that of the National Science Foundation under grant no. DMR-1714722.


\begin{thebibliography}{99}
\bibitem{smallcomment}
Full details about and suitable references to this procedure are provided elsewhere \cite{lens2}.
\bibitem{notagrsk}
In case of a positionally disordered system with a sufficiently large size,
the orientational average over the interparticle vector directions would be complete,
hence the $g(r)$ or ${\mathcal S}(k)$ curves smooth,
irrespective of whether the system be fluid or solid.
Naturally, the curve roughness also depends on the width of the bins, $\Delta r$ or $\Delta k$, of 
the histograms involved in the numerical calculation of $g(r)$ or ${\mathcal S}(k)$: 
here, $\Delta r = 0.01\sigma$ and $\Delta k \sigma = 0.1$.
\end{thebibliography}
\end{document}